\documentclass[fp,twocolumn]{jpsj3}

\usepackage{txfonts}

\usepackage{graphicx}
\usepackage{dcolumn}
\usepackage{bm}
\usepackage{color}
\usepackage{mathrsfs}

\title{ 
Statistical Properties of Current Noise Induced by Electron--Phonon Scattering in Metallic Carbon Nanotubes}
\author{Aina Sumiyoshi$^1$, Keisuke Ishizeki$^2$ and Takahiro Yamamoto$^{1,3,\dagger}$}

\inst{$^1$ Department of Physics, Tokyo University of Science, 1-3 Kagurazaka, Shinjuku, Tokyo 162-8601, Japan \\
$^2$ Department of Electrical Engineering, Tokyo University of Science, 6-3-1 Niijuku, Katsushika, Tokyo 125-8585, Japan\\
$^3$ RIST, Tokyo University of Science, 1-3 Kagurazaka Shinjuku, Tokyo 162-8601, Japan} 

\abst{
We theoretically investigate current noise in metallic carbon nanotubes induced by electron--phonon scattering, focusing on 
the probability density function (PDF) of the current that characterizes the nonequilibrium steady state. Quantum transport simulations 
combined with analyses of higher-order statistical moments reveal that the PDF evolves continuously from a Gaussian distribution in 
the ballistic regime to a non-Gaussian gamma distribution in the diffusive regime. In the crossover regime, the PDF exhibits pronounced 
asymmetry, attributed to a statistical imbalance in the number of conduction pathways contributing to high- and low-current events. 
Furthermore, in the diffusive regime, we identify non-Markovian features arising from high-frequency resonances in the current noise, 
which dominate the asymptotic scaling behavior of the current variance.
}

\begin{document}
\maketitle

\section{Introduction~\label{sec:1}}                                                          
Electronic transport phenomena at the nanoscale are strongly influenced by quantum coherence of electrons and many-body interactions 
(electron--electron, electron--phonon, etc.), making them a central focus in both mesoscopic physics and nanoelectronics~\cite{rf:datta,rf:imry}. 
Recently, current noise has emerged as an effective probe for revealing features of quantum transport at the mesoscale 
that are inaccessible through average current measurements~\cite{rf:Jong,rf:Blanter,rf:Martin,rf:Piatrusha,rf:kobayashi}. 
It serves as a sensitive indicator of the statistical properties of charge carriers, scattering processes, and the presence of 
quantum coherence. Among these, electron--phonon (e--ph) scattering plays a critical role as a primary source of phase relaxation and 
is essential for determining current noise characteristics, as it induces dynamical fluctuations of carriers. In fact, our recent theoretical study 
clarifies that e--ph scattering in carbon nanotubes (CNTs), which are realistic one-dimensional (1D) nanomaterials~\cite{rf:iijima,rf:saito,rf:hamada}, 
has a significant influence on current noise in the mesoscopic regime (the so-called ballistic--diffusive crossover regime), where 
the tube length $L$ is comparable to the electron mean free path $L_0$~\cite{rf:sumiyoshi}.

Most studies of current noise in quantum transport have primarily focused on its intensity, particularly the power spectral density of current fluctuations~\cite{rf:Jong,rf:Blanter,rf:Martin,rf:Piatrusha,rf:kobayashi}. However, a comprehensive understanding of the statistical properties 
of current fluctuations requires analysis of the probability distribution function (PDF) of the current, as well as higher-order statistical moments 
such as skewness and kurtosis~\cite{rf:papoulis}. While such descriptors have been experimentally probed in systems like tunnel junctions~\cite{rf:Reulet} 
and quantum dots~\cite{rf:Gustavsson,rf:Fujisawa}, their measurement remains challenging in many mesoscopic systems, including CNTs. 
From a theoretical standpoint, extending such analyses to realistic nanomaterials beyond simplified models remains a considerable challenge.

In this study, prior to experimental implementation, we theoretically investigate the PDF of current in CNTs  arising from e--ph scattering 
and evaluate key statistical descriptors, namely, the mean, variance, skewness, and kurtosis. This analysis is conducted using 
a quantum transport simulation framework based on the OpenTDSE+MD method, which we have recently 
developed~\cite{rf:ishizeki2017,rf:ishizeki2018,rf:ishizeki2020,rf:sumiyoshi}. This method enables the simulation of time-resolved quantum transport 
in realistic nanostructures by combining electronic dynamics (via the time-dependent Schr{\"o}dinger equation (TDSE) ) with thermal lattice vibrations 
(via molecular dynamics (MD) ). This coupling allows phonon-induced fluctuations to be naturally included in the transport calculations.

\section{Model and Methods}
\subsection{Basics of statistical analysis for current noise}
Here, we briefly summarize
the moment expansion method, a mathematical technique for extracting characteristic features from the current PDF, $P(J)$.
Moments are fundamental statistical quantities that quantitatively characterize the shape of $P(J)$. The $k$-th moment of the current $J$ 
(a random variable) is defined as
\begin{equation}
\mu_k \equiv \langle J^k \rangle = \int_0^\infty\!\!\! J^k P(J)\, dJ,
\end{equation}
where $\mu_k$ is the $k$-th raw moment about the origin, $J = 0$. 
The first four moments correspond to widely used statistical descriptors: the mean, variance, skewness, and kurtosis. The first moment, 
$\mu_1 = \langle J \rangle$, gives the mean (i.e., the average current). The second moment is related to the variance, $\sigma^2 = \mu_2 - \mu_1^2$, 
which quantifies the magnitude of current fluctuations and is directly connected to the noise power.

Higher-order moments capture more subtle characteristics of the distribution. The third moment determines the skewness,
\begin{equation}
\gamma_1 = \frac{\mu_3 - 3\mu_1\mu_2 + 2\mu_1^3}{\sigma^3},
\label{eq:gamma1}
\end{equation}
which measures the asymmetry of $P(J)$ around its mean. The fourth moment determines the kurtosis,
\begin{equation}
\gamma_2 = \frac{\mu_4 - 4\mu_1\mu_3 - 3\mu_2^2 + 12\mu_1^2\mu_2 - 6\mu_1^4}{\sigma^4},
\label{eq:gamma2}
\end{equation}
which quantifies the tail heaviness and peak sharpness relative to a Gaussian distribution.
Note that the above definition of kurtosis corresponds to the form based on raw moments, not the excess kurtosis (i.e., $\gamma_2 - 3$). 

For a purely Gaussian distribution, the skewness and kurtosis take the values $\gamma_1 = 0$ and $\gamma_2 = 3$, respectively. 
Deviations from these values therefore indicate non-Gaussian features of the current PDF, such as asymmetry and contributions from rare events.

\begin{figure}[t]
\begin{center}
\includegraphics[width=85mm]{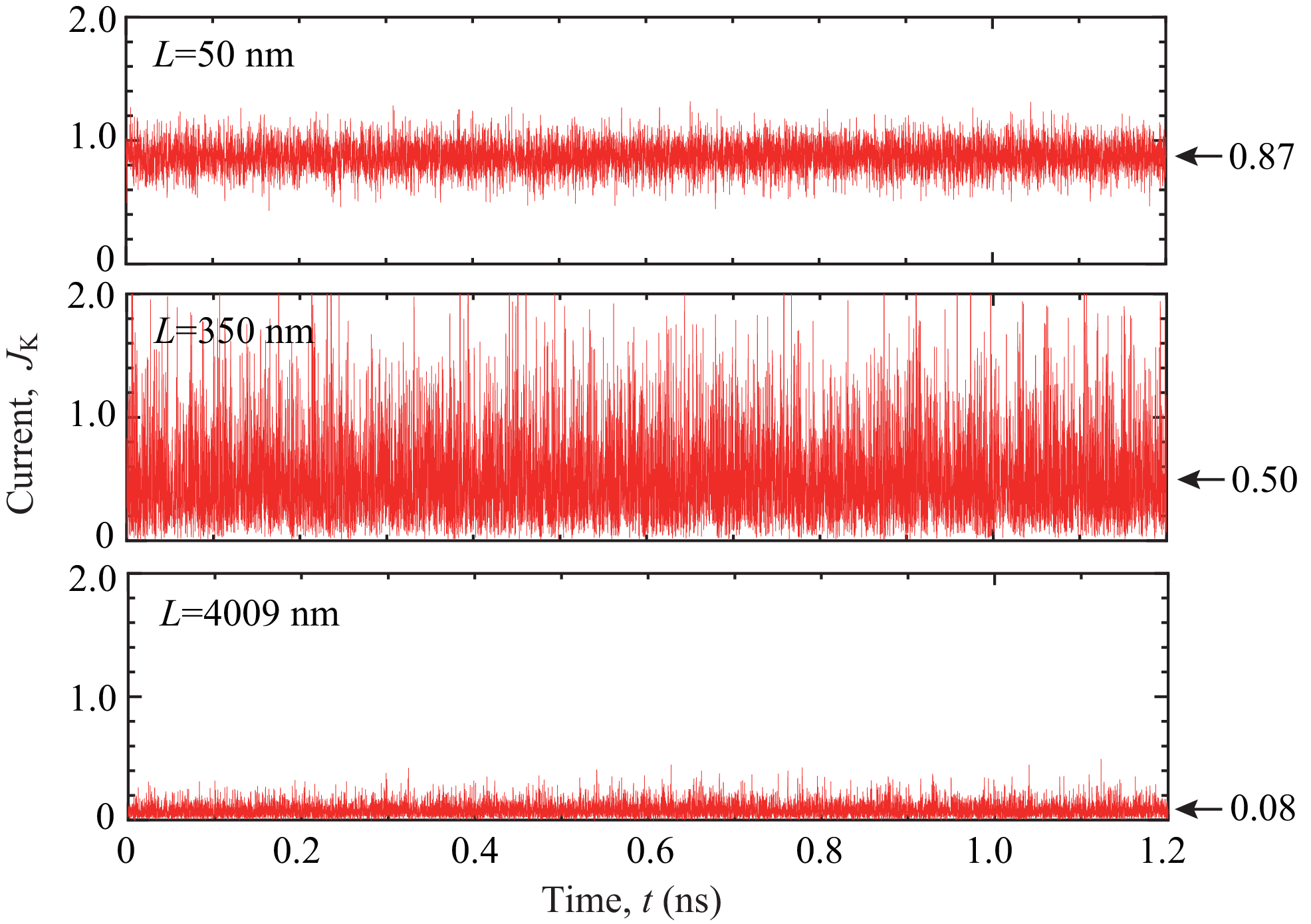}
\caption{(Color online) Time evolution of the dimensionless current $J_K(t)$ in (5,5) SWCNTs with lengths $L = 50$, 350, and 
4009~nm at 300~K. The current is normalized by $I_0 = 2e^2V/h$. The time-averaged current decreases with 
increasing tube length, while the current fluctuations exhibit a maximum around $L \approx L_0$, corresponding to the ballistic--diffusive 
crossover regime.}
\label{fig:01}
\end{center}
\end{figure}

\section{Simulation setup}
To evaluate the current through a 1D conductor of arbitrary length $L$ with e--ph interactions, we employ a two-terminal setup in which
 the finite-length conductor is connected to two ideal, non-interacting leads of semi-infinite extent. Scattering is taken into account only within the conductor. 
In this configuration, electrons injected from one lead propagate through the conductor while being scattered by phonons, and are eventually transmitted 
to the opposite lead. By counting the number of transmitted electrons as a function of time, the time-dependent current can be determined.
In this study, to eliminate contact resistance between the central region and the leads, all regions are composed of carbon nanotubes with identical chirality.

A powerful framework for simulating quantum transport in this setup is the OpenTDSE+MD method, which has been developed by our 
group~\cite{rf:ishizeki2017,rf:ishizeki2018,rf:ishizeki2020,rf:sumiyoshi}.  While the full details of the method are provided in 
Refs.~[\citen{rf:ishizeki2017,rf:ishizeki2018,rf:ishizeki2020,rf:sumiyoshi}], we briefly summarize its key features below and describe 
the specific simulation options and parameter values used in the present study. In the OpenTDSE+MD method, the nuclear motion in 
the conductor at finite temperature is simulated using MD method under the NTV ensemble~\cite{rf:Woodcoc} with velocity 
scaling~\cite{rf:Haile,rf:Anderson}. To model the interatomic forces between carbon atoms in the CNT, we adopt the Tersoff potential as 
a reliable force field~\cite{rf:lindsay}. The effect of nuclear motion is incorporated into the time-dependent hopping integrals $\gamma_{ij}(t)$ of 
the tight-binding Hamiltonian, based on the Harrison rule~\cite{rf:harrison}:
\begin{equation}
\gamma_{ij}(t) = \gamma_0 \frac{|\bm{R}_{0,i} - \bm{R}_{0,j}|^2}{|\bm{R}_i(t) - \bm{R}_j(t)|^2},
\label{eq:hopping}
\end{equation}
where $\gamma_0 = -2.7$~eV is the hopping integral at equilibrium, $\bm{R}_{0,i}$ is the equilibrium position of the $i$th carbon atom, and 
$\bm{R}_i(t)$ is its position at time $t$, obtained from the MD simulation. By solving the TDSE under the two-lead configuration using the hopping 
integrals defined in Eq.~(\ref{eq:hopping}), we obtain the time-dependent wavefunction, which is subsequently used to calculate the instantaneous 
current through the conductor.

\begin{figure}
\begin{center}
\includegraphics[width=80mm]{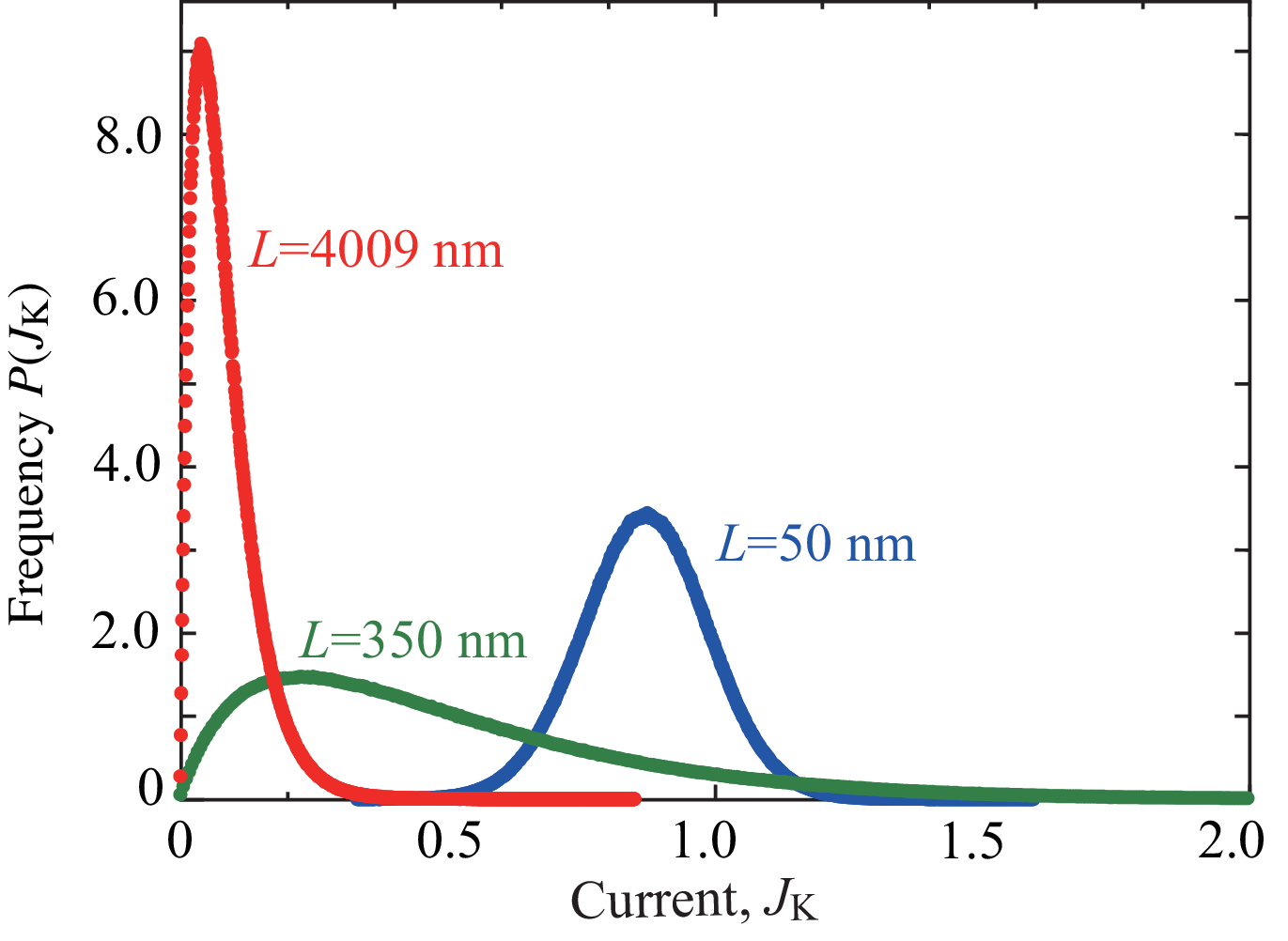}
\caption{(Color online) Probability density function (PDF) of the dimensionless current $J_{\rm K}$ in (5,5) SWCNTs of lengths $L = 50$, 350, and 
4009~nm at 300~K. In the ballistic regime ($L = 50$~nm), the PDF is well approximated by a Gaussian function. As the system transitions to 
the diffusive regime ($L = 4009$~nm), the PDF deviates from Gaussian behavior and is better described by a Gamma distribution with 
asymmetric features. The crossover regime ($L = 350$~nm) exhibits significant non-Gaussian fluctuations.}
\label{fig:02}
\end{center}
\end{figure}

In the following section, we present the phonon-induced current noise in metallic (5,5) single-walled carbon nanotubes (SWCNTs) of 
various lengths, calculated using the Open-TDSE+MD method. The length $L$ is defined as $L = n_{\mathrm{uc}} a$, where $a = 0.25$~nm is 
the unit cell length of the (5,5) SWCNT and $n_{\mathrm{uc}}$ is the number of unit cells in the scattering region. The Fermi energy is set at 
the charge neutrality point (CNP), such that the current is carried by electrons near the two CNPs. We focus on electrons near the CNP, 
also referred to as the K and K$'$ points in analogy with Dirac points in graphene, which represent the two degenerate valleys in the electronic 
band structure of (5,5) SWCNTs. Thus, the dimensionless current $J(t)$ is defined as
\begin{equation}
J(t) \equiv \frac{I(t)}{|I_0|} = J_{\mathrm{K}}(t) + J_{\mathrm{K'}}(t),
\label{eq:dimless_current}
\end{equation}
where $I(t)$ denotes the time-dependent current through (5,5) SWCNTs, and $I_0 \equiv 2e^2V/h$ is the current through a single perfectly transmitting 
channel ($e$ is the elementary charge, $h$ is Planck constant, and $V$ is the bias voltage applied between the leads). Here, $J_{\mathrm{K}}(t)$ 
and $J_{\mathrm{K'}}(t)$ are the dimensionless currents carried by electrons injected from the K and K$'$ points, respectively, which contribute equally 
to the total current. Henceforth, we focus our analysis on $J_{\mathrm{K}}(t)$ without loss of generality. In this study, we define $t = 0$ as 
a moment within the steady-state regime, i.e., after the system has evolved sufficiently for a stable current to be established in the central region. 
A time step of $\Delta t = 0.1$~fs is used throughout the simulations. The measurement time $\tau$ of current is taken to be 1.2 ns, 
which is longer than the mean free time $\tau_0 = 0.3$~ps.

\section{Simulation Results and Discussion}
\subsection{Time-dependent current and current PDF}
Figure~\ref{fig:01} shows the time-dependent current $J_{\mathrm{K}}(t)$ for (5,5) SWCNTs with $L = 50$, 350, and 4009~nm at room temperature 
($T = 300$K). As indicated by the arrows in Fig.\ref{fig:01}, the time-averaged current decreases monotonically with increasing $L$. In contrast, the 
current fluctuations around the average exhibit a non-monotonic dependence on $L$, peaking at $L = 350$~nm, which is approximately equal to the 
mean free path ($L_0 = 344$nm), as will be discussed later. In the following, instead of analyzing the time average and temporal fluctuations of the 
current, we perform a statistical analysis based on the current PDF, obtained from the time-dependent current data shown in Fig.\ref{fig:01}.

Figure~\ref{fig:02} presents the current PDF for (5,5) SWCNTs with lengths $L = 50$, 350, and 4009~nm at $T = 300$~K. The current PDF, 
$P(J_{\mathrm{K}})$, is well fitted by a Gaussian distribution in the ballistic regime ($L = 50$~nm). In contrast, in the crossover ($L = 350$~nm) 
and diffusive ($L = 4009$~nm) regimes, the distribution clearly deviates from Gaussian behavior, reflecting the presence of nontrivial current 
fluctuations beyond simple thermal noise, i.e., Johnson-Nyquist noise~\cite{rf:J, rf:N}. In the following sections, we elucidate how the current PDF 
evolves from a Gaussian to a non-Gaussian form as the system transitions from the ballistic to the diffusive regimes. This evolution is characterized 
through the analysis of statistical descriptors such as mean, variance, skewness, and kurtosis of current. Particular attention is given to the non-Gaussian 
features that emerge in the diffusive limit, where e--ph interactions become dominant.

\begin{figure}
\begin{center}
\includegraphics[width=80mm]{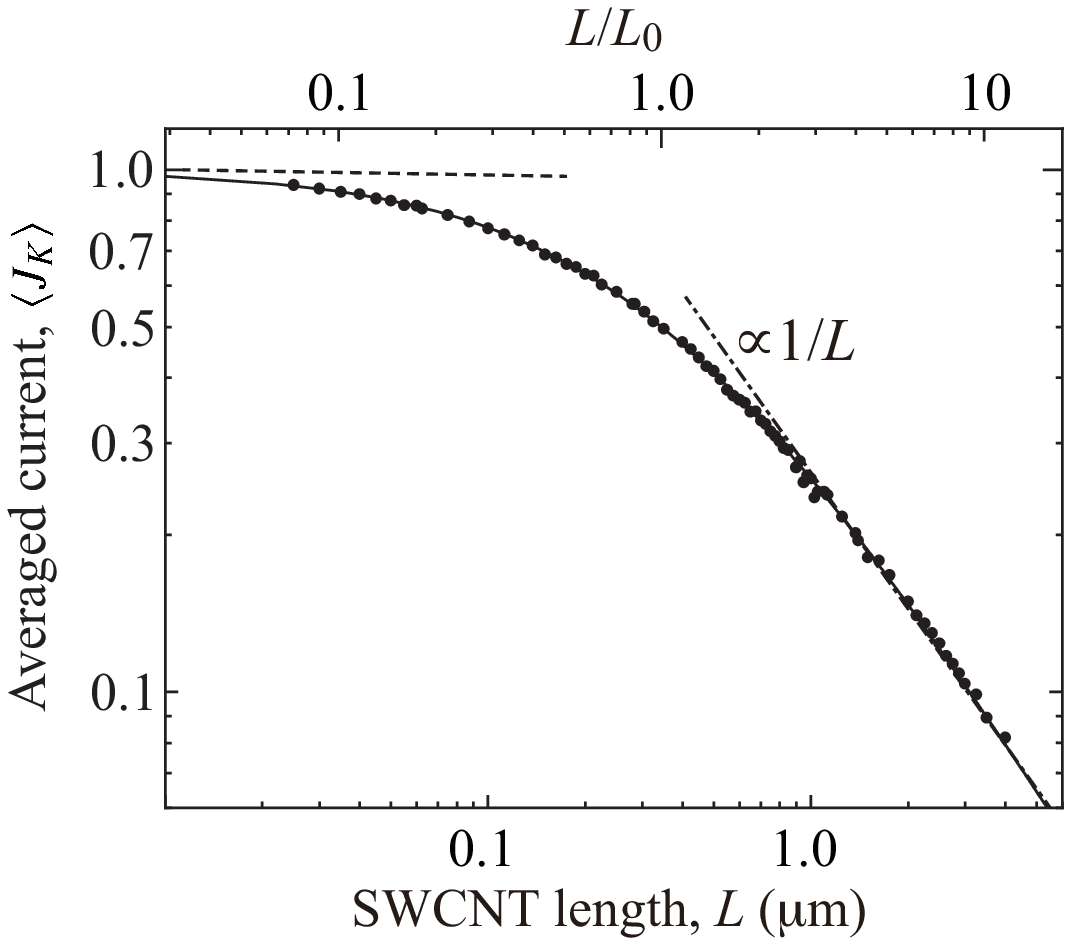}
\caption{Length dependence of the ensemble-averaged current, $\langle J_{\rm K}\rangle$, in (5,5) SWCNTs at 300 K. The solid curve 
shows the theoretical prediction, $\langle J_{\rm K}\rangle = L_0/(L_0+L)$. The horizontal dashed line indicates the averaged current in 
the diffusive limit, where $\langle J_{\rm K}\rangle=1$. The broken line represents the asymptotic Ohmic behavior in the diffusive regime, 
in which $\langle J_{\rm K}\rangle \propto 1/L$.}
\label{fig:03}
\end{center}
\end{figure}

\subsection{Mean and variance of current}
The ensemble-averaged current (i.e., mean current) is evaluated as the first-order moment $\mu_1$ of $J_{\mathrm{K}}$:
\begin{equation}
\langle J_{\mathrm{K}} \rangle = \mu_1 \equiv \int_0^\infty J_{\mathrm{K}} P(J_{\mathrm{K}})~dJ_{\mathrm{K}},
\end{equation}
where $P(J_{\mathrm{K}})$ is the current PDF shown in Fig.\ref{fig:02}.
Figure~\ref{fig:03} presents the $L$ dependence of $\langle J_{\mathrm{K}} \rangle$ for the (5,5) SWCNT at $T = 300$~K. 
In the ballistic limit ($L/L_0 \to 0$), $\langle J_{\mathrm{K}} \rangle$ approaches unity, indicating that the electrical conductance is quantized at 
the universal value of $2e^2/h$. In contrast, in the diffusive regime ($L/L_0 \gg 1$), $\langle J_{\mathrm{K}} \rangle$ decreases inversely with $L$, 
as shown by the dashed line in Fig.\ref{fig:03}, in agreement with Ohm's law.

\begin{figure}
\begin{center}
\includegraphics[width=80mm]{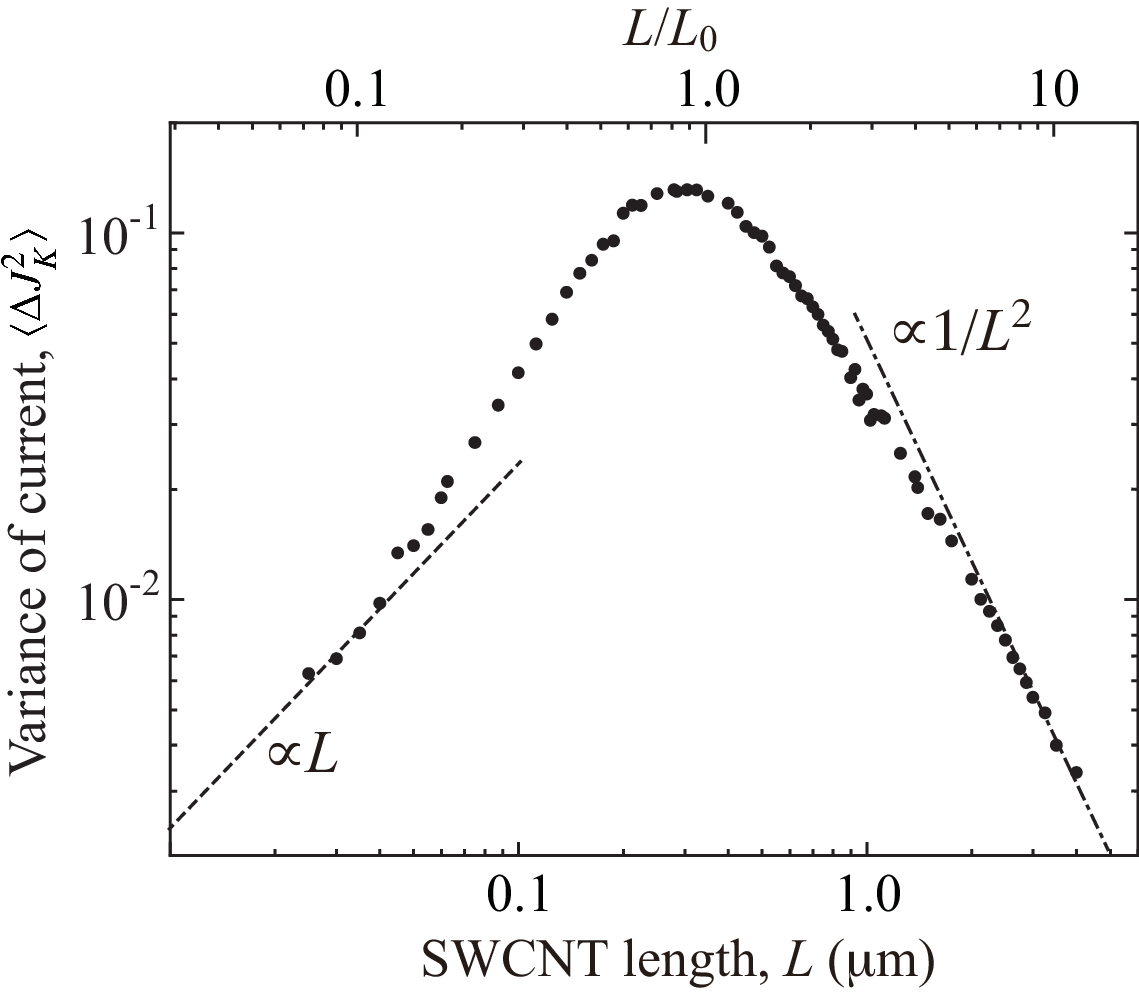}
\caption{Length dependence of the current variance $\langle \Delta J_K^2 \rangle$ in (5,5) SWCNTs at 300\,K. The variance exhibits 
a non-monotonic dependence on the tube length, peaking around $L/L_0 \approx 1$ due to maximum uncertainty in transmission and 
reflection probabilities. In the ballistic regime ($L/L_0 \ll 1$), the variance increases linearly with length, whereas, in the diffusive regime 
$L/L_0\gg 1$, it appears to follow a power-law decay $L^{-\alpha}$ and the exponent is $\alpha =2$.}
\label{fig:04}
\end{center}
\end{figure}

As discussed in our previous studies~\cite{rf:ishizeki2017,rf:sumiyoshi}, the crossover from ballistic to diffusive transport can be understood in terms of 
an effective transmission function ${\mathcal T}_{\mathrm{eff}}$ that incorporates phase-breaking effects due to e--ph scattering~\cite{rf:datta}:
\begin{equation}
{\mathcal T}_{\mathrm{eff}} = \frac{L_0}{L_0 + L},
\label{eq:transmission_function}
\end{equation}
which is equivalent to the average of the dimensionless current, i.e., $\langle J_{\mathrm{K}} \rangle={\mathcal T}_{\rm eff}$.
The simulation data in Fig.\ref{fig:03} show excellent agreement with the theoretical prediction given by Eq.~(\ref{eq:transmission_function}) with 
$L_0 = 344$~nm, validating the model's ability to capture inelastic transport across a wide length scale from the ballistic to the diffusive regime. 
Note that the \textit{ensemble}-averaged current in Fig.~\ref{fig:03} is in good agreement with the \textit{time}-averaged current reported in 
our previous studies~\cite{rf:ishizeki2017,rf:sumiyoshi}.

We next evaluate the current variance of (5,5) SWCNTs at $T = 300$~K. By calculating the second moment $\mu_2$ defined as
\begin{equation}
\mu_2 = \int_0^\infty J_{\mathrm{K}}^2 P(J_{\mathrm{K}})~dJ_{\mathrm{K}},
\end{equation}
we obtain the current variance $\langle \Delta J_{\mathrm{K}}^2 \rangle =\sigma^2\equiv \mu_2 - \mu_1^2$, 
where $\Delta J_{\mathrm{K}} = J_{\mathrm{K}}(t) - \langle J_{\mathrm{K}} \rangle$ denotes the deviation from the averaged current.
Figure~\ref{fig:04} shows the length dependence of the current variance for (5,5) SWCNTs at $T = 300$~K. In contrast to the monotonic 
decrease of the averaged current in Fig.~\ref{fig:03}, the current variance exhibits a non-monotonic dependence on $L$, peaking around 
$L/L_0 \approx 1$. This result is physically intuitive: at this length scale, the probability of an electron being transmitted through the SWCNT 
without scattering becomes comparable to the probability of being reflected. This balance leads to a maximum uncertainty in the instantaneous current.

We see from the numerical data in Fig.~\ref{fig:04} that $\langle \Delta J_{\mathrm{K}}^2 \rangle$ increases linearly with $L$ in 
the ballistic transport regime ($L/L_0 \ll 1$), whereas it appears to follow a power-law decay $L^{-\alpha}$ with $\alpha \approx 1.7$ in 
the diffusive transport regime ($L/L_0 \gg 1$). A detailed physical explanation of these asymptotic behaviors of 
$\langle \Delta J_{\mathrm{K}}^2 \rangle$ will be provided in Sec.~5.

\subsection{Skewness and kurtosis of current}

\begin{figure}
\begin{center}
\includegraphics[width=80mm]{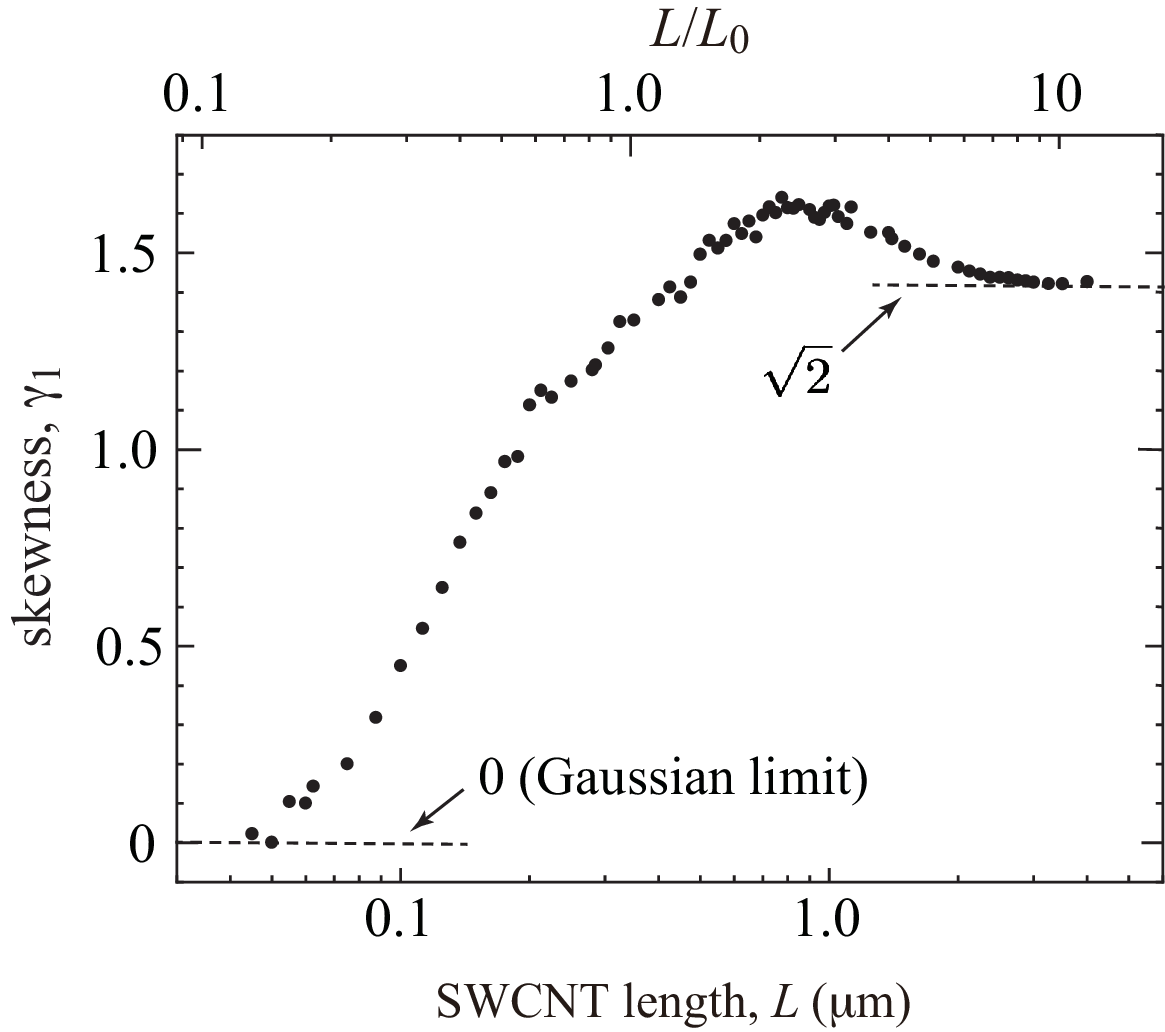}
\caption{Length dependence of the skewness $\gamma_1$ of the current in (5,5) SWCNTs at 300~K. The skewness is nearly zero in 
the ballistic regime ($L/L_0 \ll 1$), indicating a symmetric (Gaussian) current distribution. As the tube length increases, $\gamma_1$ 
grows and peaks around $L/L_0 \approx 2$, reflecting the emergence of asymmetric fluctuations in the crossover regime. In the diffusive limit 
($L/L_0 \gg 1$), the skewness saturates to $\sqrt{2}$, consistent with a Gamma distribution with shape parameter $n = 2$.}
\label{fig:05}
\end{center}
\end{figure}

To comprehensively characterize the shape of the current distribution, we analyze its statistical descriptors, including skewness and kurtosis.
The third moment $\mu_3$ is calculated as
\begin{equation}
\mu_3 = \int_0^\infty J_{\mathrm{K}}^3 P(J_{\mathrm{K}})~dJ_{\mathrm{K}},
\end{equation}
and, using the standard deviation $\sigma$ along with the first and second moments $\mu_1$ and $\mu_2$, the skewness $\gamma_1$, 
which quantifies the asymmetry of the distribution, is evaluated. The skewness is defined by Eq.~(\ref{eq:gamma1}). Figure~\ref{fig:05} shows 
the length dependence of the current skewness for (5,5) SWCNTs at room temperature ($T = 300$~K). In the ballistic limit ($L/L_0 \to 0$), 
$\gamma_1$ approaches zero, consistent with the symmetric nature of a Gaussian distribution. As $L$ increases, $\gamma_1$ gradually 
becomes finite, exhibits a peak near $L/L_0 \approx 2$, and eventually approaches $\sqrt{2}$ in the diffusive limit ($L/L_0 \gg 1$), indicating 
a clear deviation from Gaussian statistics (the origin of the limiting value $\gamma_1 = \sqrt{2}$ will be discussed later). A positive value of 
$\gamma_1$ indicates that $P(J_{\mathrm{K}})$ is skewed to the right, with a longer tail on the high-current side. In other word, 
this asymmetry of PDF arises from a statistical imbalance in the number of conduction paths contributing to low- and high-current states.

\begin{figure}
\begin{center}
\includegraphics[width=80mm]{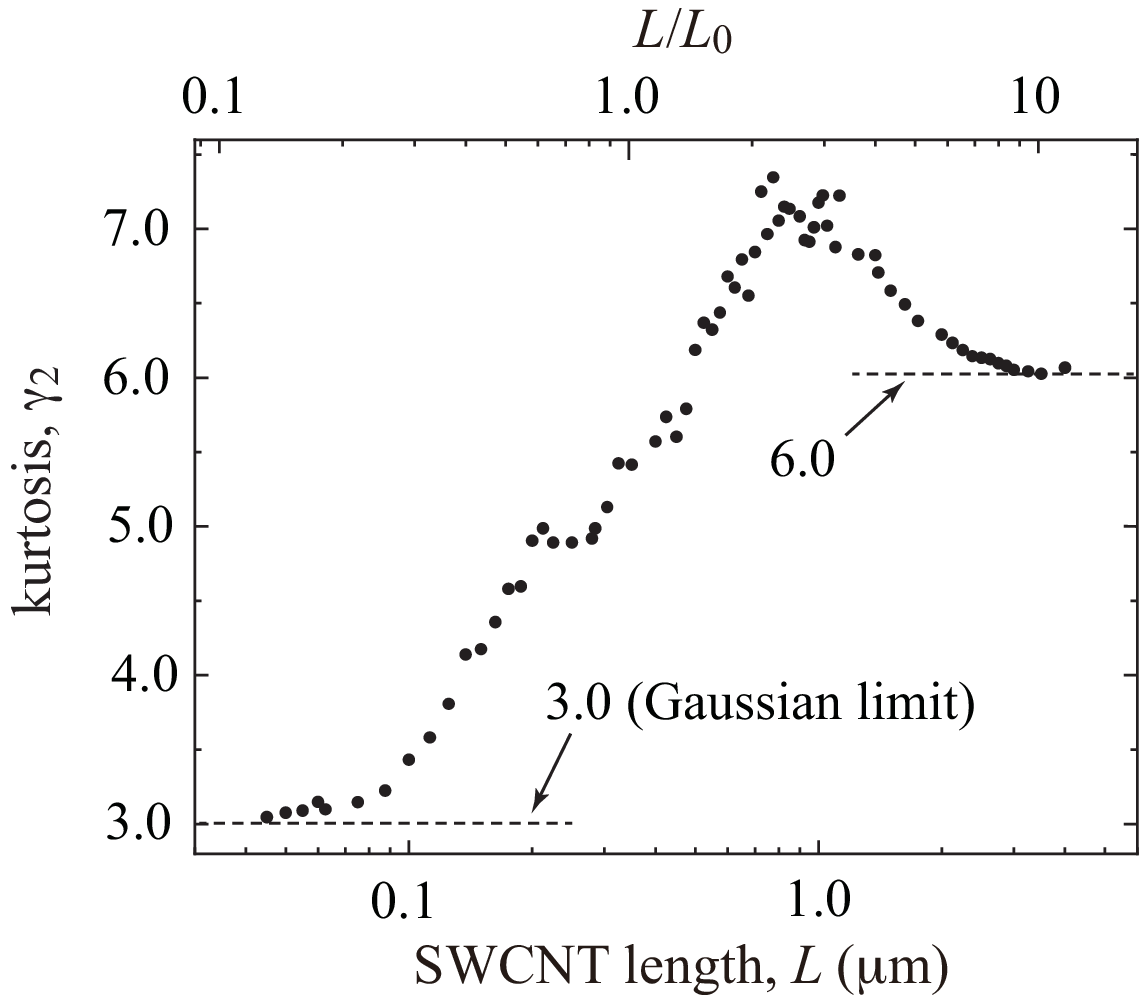}
\caption{Length dependence of the kurtosis $\gamma_2$ of the current in (5,5) SWCNTs at 300~K. In the ballistic regime ($L/L_0 \ll 1$), 
the kurtosis approaches 3, consistent with a Gaussian distribution. As the length increases, $\gamma_2$ rises, exhibiting a peak and 
eventually saturating at 6 in the diffusive regime ($L/L_0 \gg 1$). These findings indicate a transition to a heavy-tailed current distribution 
described by a Gamma distribution with shape parameter $n = 2$.}
\label{fig:06}
\end{center}
\end{figure}

Next, we consider the kurtosis, which quantifies the peakedness (or flatness) of the distribution. The fourth moment $\mu_4$ is given by
\begin{equation}
\mu_4 = \int_0^\infty J_{\mathrm{K}}^4 P(J_{\mathrm{K}})~dJ_{\mathrm{K}},
\end{equation}
and the kurtosis $\gamma_2$ is defined by Eq.~(\ref{eq:gamma2}).
Figure~\ref{fig:06} presents the length dependence of the current kurtosis for (5,5) SWCNTs at $T = 300$~K. In the ballistic limit ($L/L_0 \to 0$), 
$\gamma_2$ approaches 3, consistent with the kurtosis of a Gaussian distribution. As $L$ increases, $\gamma_2$ gradually rises, exhibits a peak, 
and eventually approaches 6 in the diffusive limit ($L/L_0 \gg 1$), further highlighting a significant deviation from Gaussian behavior. A value of 
$\gamma_2 > 3$ implies that the distribution has a sharper peak and heavier tails than a Gaussian distribution. In other words, the distribution is 
more tightly concentrated around the mean while also exhibiting a higher probability of extreme fluctuations.

We now turn to a consideration of the statistical distribution that gives rise to $\gamma_1 = \sqrt{2}$ and $\gamma_2 = 6$. Based on these 
two values, our results clearly demonstrate that the PDF converges to a gamma distribution with shape parameter $n = 2$, defined as
\begin{equation}
P(J_{\mathrm{K}}) = \frac{\xi^n}{\Gamma(n)} J_{\mathrm{K}}^{n-1} e^{-\xi J_{\mathrm{K}}},
\label{eq:gamma}
\end{equation}
where $\xi$ is the rate parameter and $\Gamma(n)$ is the gamma function with integer shape parameter $n$. From this expression, 
standard statistical descriptors such as the mean, variance, skewness, and kurtosis are readily obtained as
\begin{equation}
\langle J_{\rm K} \rangle = \frac{n}{\xi}, \quad
\sigma^2 = \frac{n}{\xi^2}, \quad
\gamma_1 = \frac{2}{\sqrt{n}}, \quad
\gamma_2 = 3 + \frac{6}{n}.
\label{eq:SD_gamma}
\end{equation}
For \( n = 2 \), Eq.~(\ref{eq:SD_gamma}) yields \( \gamma_1 = \sqrt{2} \) and \( \gamma_2 = 6 \), which are consistent with the limiting values 
observed in the diffusive regime. 

In the following section, we explain why the current PDF of SWCNTs approaches a Gaussian distribution in the ballistic regime and a Gamma 
distribution with shape parameter $n=2$ in the diffusive regime, by proposing stochastic models of electron transport in each respective limit.

\section{Physical Interpretation for Current PDFs}
To complement the numerical results and deepen physical understanding, we present simplified stochastic models that qualitatively reproduce 
the observed current PDFs in both the ballistic and diffusive regimes.

\subsection{Current PDF in the ballistic regime ($L/L_0 \ll 1$)}
In this section, we formulate the current PDF in the ballistic transport regime ($L/L_0 \ll 1$) based on a statistical approach. For notational simplicity, 
we denote $J_{\mathrm{K}}$ simply as $J$ in the following.

We consider a one-dimensional conductor of length $L$ connected to two ideal leads (i.e., the Landauer-type model), where electrons are 
scattered solely by electron--phonon (e--ph) interactions within the conductor. In this study, we focus on the \textit{dephasing effect} induced by 
e--ph scattering. The scattering is treated as \textit{elastic}, and its influence is characterized by the \textit{effective transmission probability} 
$\mathcal{T}_{\rm eff} \equiv \mathcal{T}_{\rm eff}(\varepsilon_{\mathrm{F}})$, which is valid in the limit $\mathcal{T}_{\mathrm{eff}} \approx 1$. 
The replacement of e--ph interactions with an effective transmission probability is a well-established concept in quantum transport theory and 
has been systematically employed in frameworks such as the nonequilibrium Green's function (NEGF) method and B{\" u}ttiker's fictitious 
probe model~\cite{rf:datta}.

The applied voltage between the electrodes is assumed to be sufficiently small such that only electrons at the Fermi energy $\varepsilon_{\mathrm{F}}$ 
contribute to transport. Under this condition, the probability $P(n)$ that $n$ out of $N$ incident electrons per unit time are transmitted through 
the conductor is given by a binomial distribution:
\begin{equation}
P(n) = {}_N C_n~ \mathcal{T}_{\rm eff}^n (1 - \mathcal{T}_{\mathrm{eff}})^{N - n},
\label{eq:binomial}
\end{equation}
where ${}_N C_n = \dfrac{N!}{n!(N - n)!}$ is the binomial coefficient, and $\mathcal{T}_{\rm eff}$ is evaluated at the Fermi energy.

Since the dimensionless current $J$ is equivalent to $n/N$, Eq.~\eqref{eq:binomial} can be rewritten as $P(J)$. 
In the limit in which $n/N$ can be regarded as a continuous variable and $T_{\mathrm{eff}} \approx 1$, the discrete binomial 
distribution converges to a continuous Gaussian distribution:
\begin{equation}
P(J) = \frac{1}{\sqrt{2\pi \sigma^2}} \exp\left(-\frac{(J - \langle J \rangle)^2}{2\sigma^2}\right),
\end{equation}
where $\langle J \rangle$ and $\sigma^2 \equiv \langle \Delta J^2 \rangle$ are the mean and variance of the current $J$, respectively, given by
\begin{eqnarray}
\langle J \rangle &=&  \mathcal{T}_{\rm eff}, \label{eq:mean} \\
\langle \Delta J^2 \rangle &=& \frac{\mathcal{T}_{\rm eff} (1 - \mathcal{T}_{\rm eff})}{N}
\approx \frac{(1 - \mathcal{T}_{\rm eff})}{N}.
 \label{eq:variance}
\end{eqnarray}
Thus, in the ballistic regime ($L/L_0 \ll 1$), the skewness and kurtosis of the current distribution are $\gamma_1 = 0$ and $\gamma_2 = 3$, respectively.

Based on this scenario, we can explain the linear dependence of the variance $\langle \Delta J^2 \rangle$ on $L$ in the ballistic regime ($L/L_0 \ll 1$), 
as shown in Fig.~\ref{fig:04}. By substituting Eq.~(\ref{eq:transmission_function}) into Eq.~(\ref{eq:variance}) and taking the limit $L/L_0 \to 0$, 
we readily obtain $\langle \Delta J^2 \rangle \propto L/L_0$.

\subsection{Current PDF in diffusive regime ($L/L_0\gg 1$)}
In this section, we develop a stochastic model to describe the current PDF of armchair-type SWCNTs in the diffusive transport regime, 
where $L/L_0 \gg 1$. We demonstrate that the current PDF in this limit follows a Gamma distribution with shape parameter $n = 2$. 
We consider an electron injected from the left lead into the central scattering region in either the K or K$'$ valley state. During its propagation, 
the electron undergoes frequent inter-valley scattering events (K $\leftrightarrow$ K$'$) mediated by e--ph interactions. The valley dynamics are 
modeled as a symmetric two-state Markov process between the K and K$'$ states. The electron is eventually transmitted through the scattering region 
while occupying either valley state, and we define this moment as time $t' = 0$ for convenience in the following discussion.

We define the survival probability $S_{\mathrm{K(K')}}(t)$ as the probability that the system remains in the K (or K$'$) valley state without flipping 
during the time interval from $t' = -t$ to $t' = 0$. Now consider a slightly earlier time $t' = -(t + \Delta t)$. The probability that the electron has remained 
in the K state during the extended interval from $t' = -(t + \Delta t)$ to $t' = 0$ is given by
\begin{equation}
S_{\mathrm{K}}(t + \Delta t) = S_{\mathrm{K}}(t) \left(1 - \frac{\Delta t}{\tau} \right),
\label{eq:A_S(t+deltat)}
\end{equation}
where $\Delta t \ll \tau$, and $\tau$ denotes the mean time between valley-flipping events. Here, $\Delta t / \tau$ represents the probability that 
a valley flip occurs during the infinitesimal time interval $\Delta t$. Taking the limit $\Delta t \to 0$ in Eq.~(\ref{eq:A_S(t+deltat)}), we obtain 
the following first-order differential equation:
\begin{equation}
\frac{dS_{\mathrm{K}}(t)}{dt} = -\frac{1}{\tau} S_{\mathrm{K}}(t).
\label{eq:dS/dt}
\end{equation}
Solving Eq.~(\ref{eq:dS/dt}) with the initial condition $S_{\rm K}(0) = 1$, we find the survival probability takes an exponential form:
\begin{equation}
S_{\mathrm{K}}(t) = e^{-t/\tau}, \qquad t \ge 0.
\end{equation}
Furthermore, the instantaneous probability per unit time that a valley-flipping event occurs at time $t$ is given by the probability density function
\begin{equation}
f_{\mathrm{K}}(t) \equiv \frac{1}{\tau} S_{\mathrm{K}}(t) = \frac{1}{\tau} e^{-t/\tau}, \qquad t \ge 0.
\label{eq:f(t)}
\end{equation}
This exponential distribution applies symmetrically to both valley states, owing to the assumed symmetry of the inter-valley scattering process, 
i.e., $f_{\mathrm{K'}}(t) = f_{\mathrm{K}}(t)$.

Under the assumption that the electron is transmitted through the scattering region with a constant escape rate $\gamma$ while in the K state, 
the dimensionless current contribution $J^{({\rm K})}$ from the K component can be expressed as $J^{({\rm K})} = \gamma t_{\rm K}$, where 
$t_{\rm K}$ denotes the sojourn time during which the electron remained continuously in the K valley state immediately prior to transmission. 
Similarly, the transmission contribution from the K$'$ component is given by $J^{(\rm K')} = \gamma t_{\rm K'}$. Thus, both $J^{({\rm K})}$ and 
$J^{({\rm K'})}$ follow an exponential distribution $f_\alpha(J_\alpha) = \frac{1}{\gamma\tau} e^{-\frac{J_\alpha}{\gamma\tau}}$ with $\alpha=$K 
and K$'$. Consequently, the PDF of the total dimensionless current  $J= J^{({\rm K})} + J^{({\rm K'})}$ is given by the convolution of two independent 
exponential distributions:
\begin{eqnarray}
P(J) = \int_0^J f_{\rm K}(s) f_{\rm K'}(J - s) \, ds = \xi^2 J e^{-\xi J}.
\end{eqnarray}
As seen from Eq.~(\ref{eq:gamma}), this is the Gamma distribution with shape parameter $n=2$ and scale parameter $\xi^{-1}\equiv\gamma\tau$.
Here, we can obtain $\langle t_{\rm K}\rangle=\langle t_{\rm K'}\rangle=\tau$ and therefore $\langle{J}^{(\rm K)}\rangle=\langle{J}^{(\rm K')}\rangle=\gamma\tau$
and $\langle J\rangle=2\gamma\tau$. Importantly, the statistical properties of the current noise induced by phonon scattering in the diffusive regime are 
fully characterized by a single parameter $\xi$. The $k$th moment of the Gamma distribution with $n=2$ is given by
\begin{equation}
\langle J^k \rangle = \frac{(k+1)!}{\xi^k},
\end{equation}
which depends only on $k$ and $\xi$.

Finally, we discuss the $L$-dependence of $\langle \Delta J^2 \rangle$ in the diffusive regime ($L/L_0 \gg 1$), as shown in Fig.~\ref{fig:04}.
From Eq.~(\ref{eq:SD_gamma}), the variance is given by\begin{equation}
\langle \Delta J^2 \rangle = \sigma^2 = \frac{\langle J \rangle^2}{n}.
\end{equation}
Since $\langle J \rangle \approx L_0/L$ in the diffusive regime, it follows that $\langle \Delta J^2 \rangle \propto L^{-2}$, in agreement with the simulation results shown in Fig.~\ref{fig:04}. In addition, we discuss this $L^{-2}$ dependence from another perspective.
In general, the variance $\langle \Delta J^2 \rangle$ can be expressed in terms of the power spectral density (PSD) as
\begin{equation}
\langle \Delta J^2 \rangle = \frac{1}{2\pi} \int_0^\infty S(\omega) \, d\omega,
\label{eq:PSD}
\end{equation}
where $S(\omega)$ denotes the PSD of current fluctuations~\cite{rf:Jong,rf:Blanter,rf:Martin,rf:Piatrusha,rf:kobayashi,rf:raimu}.
When the electronic transport process can be modeled as a Markov process, $S(\omega)$ takes the Lorentzian form:
\begin{equation}
S(\omega, L) = \frac{S_0(L)}{1 + (\tau_{\rm c} \omega)^2},
\label{eq:Somega}
\end{equation}
where $S_0(L)$ is the zero-frequency PSD for a system of length $L$, and $\tau_{\rm c}$ is the characteristic decay time of the autocorrelation function of 
the current fluctuations. Substituting Eq.~\eqref{eq:Somega} into Eq.~\eqref{eq:PSD}, we obtain
\begin{equation}
\langle \Delta J^2 \rangle = \frac{S_0(L)}{4\tau_{\rm c}}.
\label{eq:J2_S0}
\end{equation}
In our previous study~\cite{rf:sumiyoshi}, we found that $S_0(L)$ exhibits a power-law decay as $S_0(L) \propto 1/L^{3.81}$ for (5,5) SWCNTs at 
300~K. If the assumption of Markovian dynamics holds, Eq.~(\ref{eq:J2_S0}) leads to $\langle \Delta J^2 \rangle \propto 1/L^{3.81}$. However, 
the simulation results in Fig.~\ref{fig:04} yield an exponent $\alpha =2$, which is significantly smaller than $3.81$. This discrepancy suggests that 
non-Markovian effects play a significant role in the electronic transport properties of SWCNTs.

In fact, we recently reported that $S(\omega, L)$ for SWCNTs exhibits multiple resonance peaks in the high-frequency regime, originating from
the e--ph scattering by various optical phonons~\cite{rf:raimu}, such as the RBM, oTO, and G-mode phonons. These peaks cannot be captured by 
the simple Lorentzian lineshape in Eq.~\eqref{eq:Somega}~\cite{rf:raimu} based on the Markovian process. Specifically, it can be expressed as
\begin{equation}
S(\omega, L) = \frac{S_0(L)}{1 + (\tau_{\rm c} \omega)^2} + \sum_{n=1}^{N} \frac{A_n(L)}{(\omega - \omega_n)^2 + \gamma_n(L)^2},
\label{eq:Somega_modified}
\end{equation}
where $A_n(L)$, $\omega_n$, and $\gamma_n(L)$ are the amplitude, center frequency, and linewidth of the $n$th resonance peak, respectively.
In general, $A_n(L)$ and $\gamma_n(L)$ depend on $L$, while the $L$-dependence of $\omega_n$ is relatively weak.
Taking the above result into account, we find that $\langle \Delta J^2 \rangle$ can be analytically expressed as
\begin{equation}
\langle \Delta J^2 \rangle =
\frac{S_0(L)}{4\tau_{\rm c}}
+ \frac{1}{2\pi} \sum_{n=1}^{N} \frac{A_n(L)}{\gamma_n(L)} \left[ \frac{\pi}{2} + \arctan\left( \frac{\omega_n}{\gamma_n(L)} \right) \right],
\label{eq:M-nonM}
\end{equation}
where the first term represents the Markovian contribution, and the second term arises from non-Markovian resonances in the high-frequency regime.
If the second term decays more slowly with respect to $L$ than the first term, it governs the asymptotic behavior of $S(\omega, L)$. In this context, 
we consider that the observed power-law behavior $S(\omega, L) \propto L^{-\alpha}$ with $\alpha =2$ in the diffusive regime (as shown in 
Fig.~\ref{fig:04}) originates from the non-Markovian resonances. That is, the second term in Eq.~\eqref{eq:M-nonM} follows a power-law decay with 
respect to $L$ as $L^{-\alpha}$. To theoretically derive this behavior, one would need to determine the detailed $L$-dependence of both $A_n(L)$ and 
$\gamma_n(L)$. However, such an analysis lies beyond the scope of the present study and is left for future work.

\section{Conclusion}
We have theoretically investigated the statistical properties of current noise in metallic SWCNTs under the influence of e--ph scattering, focusing 
particularly on the evolution of the current PDF across different transport regimes. By employing a quantum transport simulation framework based on 
the OpenTDSE+MD method, we successfully analyzed the time-dependent current and extracted statistical descriptors such as the mean, variance, 
skewness, and kurtosis. Our results demonstrate a seamless transition in the current PDF from a Gaussian distribution in the ballistic regime ($L \ll L_0$) to 
a Gamma distribution with a shape parameter of $n = 2$ in the diffusive regime ($L \gg L_0$). Notably, the skewness and kurtosis reach asymptotic values of 
$\sqrt{2}$ and $6$, respectively, in the diffusive limit, indicating significant non-Gaussian features arising from phonon-induced fluctuations. Furthermore, 
we developed intuitive stochastic models to physically interpret the origins of the observed PDFs in each regime. In the ballistic regime, the current statistics 
can be described by a binomial process that approaches Gaussian behavior in the limit of high transmission probability. In contrast, in the diffusive regime, 
the current distribution emerges from the convolution of independent exponential processes associated with valley dynamics, leading to a Gamma distribution.

In addition, we examined the asymptotic behavior of the current variance $\langle \Delta J^2 \rangle$ in the diffusive regime ($L \gg L_0$) and found that 
it follows a power-law decay with respect to $L$. Interestingly, the observed exponent $\alpha = 2$ is significantly smaller than the value predicted by 
a simple Markovian model. This discrepancy can be attributed to non-Markovian effects arising from high-frequency resonance peaks in the current power
spectral density $S(\omega, L)$, which are induced by electron--phonon interactions with various optical phonon modes. Our analysis shows that these 
non-Markovian resonances contribute an additional term to $\langle \Delta J^2 \rangle$ that decays more slowly with $L$ than the Markovian component, 
and thus dominates the asymptotic scaling in the diffusive regime.

These findings provide new insights into the fundamental nature of current fluctuations in one-dimensional nanomaterials and establish a theoretical 
foundation for interpreting nonequilibrium noise signatures in realistic nanoscale systems. They also highlight the potential of current noise as 
a diagnostic tool for distinguishing between transport regimes and for probing microscopic scattering processes.

\section*{Acknowledgements}
We would like to thank Satofumi Souma, Kenji Sasaoka and Raimu Akimoto for useful discussions regarding this research. 
This work was supported in part by the Japan Society for the Promotion of Science KAKENHI (grant no. 23H00259) and by 
Toshiba Electronic Devices \& Storage Corporation under an Academic Encouragement Grant.\\

\noindent $\dagger$ takahiro@rs.tus.ac.jp

\end{document}